\input{aipcheck}

\documentclass[
    ,final            
  ]
  {aipproc}

\layoutstyle{8x11double}

\begin{document}

\title{Solar wind turbulent spectrum from MHD to electron scales}


\author{O. Alexandrova}{
address={LESIA, Observatoire de Paris, 92190 Meudon, France.}
,altaddress={Institute of Geophysics and Meteorology, University of Cologne,  50923, Cologne, Germany.} 
}

\author{J. Saur}{
  address={Institute of Geophysics and Meteorology, University of Cologne,  50923, Cologne, Germany.}
}

\author{C. Lacombe}{
  address={LESIA, Observatoire de Paris, 92190 Meudon, France.}
}

\author{A. Mangeney}{
  address={LESIA, Observatoire de Paris, 92190 Meudon, France.}
}

\author{S. J. Schwartz}{
 address={Blackett Laboratory, Imperial College London, London SW7 2AZ, UK.}
 }
\author{J. Mitchell}{address={Blackett Laboratory, Imperial College London, London SW7 2AZ, UK.}}

\author{R. Grappin}{address={LUTH, Observatoire de Paris, 92190 Meudon, France.}} 

\author{P. Robert}{address={LPP, 10--12 avenue de l'Europe 78140 Velizy France.}}

\begin{abstract}
Turbulent spectra of magnetic fluctuations in the free solar wind are studied from MHD to electron scales using Cluster observations. We discuss the problem of the instrumental noise and its influence on the measurements at the electron scales. We confirm the presence of a curvature of the spectrum $\sim \exp{\sqrt{k\rho_e}}$ over the broad frequency range $\sim[10,100]$~Hz, indicating the presence of a dissipation. Analysis of seven spectra under different plasma conditions show clearly the presence of a quasi-universal power-law spectrum at MHD and ion scales.  However, the transition from the inertial range $\sim k^{-1.7}$ to the spectrum at ion scales $\sim k^{-2.7}$ is not universal.  Finally, we discuss the role of different kinetic plasma scales on the spectral shape, considering normalized dimensionless spectra. 
\end{abstract}

\maketitle

\section{Introduction}
Space plasmas are usually in a turbulent state, and  the solar wind is one of the closest laboratories of space plasma turbulence, where in-situ measurements are possible thanks to a number of space missions \cite{noi}.  It is  well known that at MHD scales (frequencies below $\sim 0.3$~Hz, at $1$~AU) the spectrum of magnetic fluctuations in the solar wind  has a Kolmogorov' s power law  shape $\sim f^{-5/3}$.  However, the characteristics of turbulence in the vicinity of the kinetic plasma scales (such as the inertial lengths $\lambda_{i,e} = c/\omega_{pi,e}$, $c$ being the speed of light and $\omega_{pi,e}$ the plasma frequencies of ions and electrons, respectively, the Larmor radii $\rho_{i,e}=V_{\perp i,e}/\omega_{ci,e}$ and the cyclotron frequencies $\omega_{ci,e} = eB/m_{i,e}$) are not well known experimentally and are a matter of debate. It was shown that at ion scales the turbulent spectrum has a break, and steepens to $\sim f^{-s}$, with a spectral index $s$ that is clearly non-universal, taking on values between $2$ and $4$ \citep{Leamon1998,Smith2006}. These indices were obtained  from data covering  a range of scales which did not extend very much above  the spectral break,  typically up to $\sim 1$~Hz.
 In \cite{alex09} we show that at ion scales, for a wider range of frequencies (up to $10$~Hz), magnetic spectra measured under different plasma conditions (but always for an angle between the mean magnetic field ${\bf B}$   and solar wind velocity ${\bf V}$ close to 90 degree) display a quasi-universal power-law spectrum with $s=2.8$. In the present paper  we verify this result and we clarify the difference with \citep{Smith2006}. 
 
 At electron scales, the observations are difficult and our knowledge is very poor.  The first results on the magnetic turbulence at these small scales were provided recently by the Cluster observations in the solar wind and in the Earth's magnetosheath  \cite{Mangeney2006,Lacombe2006,alexandr08angeo,Fouad09,alex09}. 
 
Solar wind observations of Sahraoui et al. \cite{Fouad09} show that at $k\rho_e\simeq k\lambda_e=1$ there exists a second spectral break with a steep  power-law $\sim f^{-4}$ at smaller scales.  Solar wind observations by Alexandrova et al. \cite{alex09} show that for $k\rho_e\sim[0.1,1]$ the spectrum is no longer  of the power law type but appears to follow an exponential dependence. At scales $k\rho_e\sim k\lambda_e>1$ the spectrum deviates from the exponential. 

Finally, we consider  normalized dimensionless spectra and we discuss the role of different kinetic plasma scales.

\section{Turbulent spectrum at electron scales}

In the solar wind plasma at 1~AU, the  electron scales are usually of the order of a few km, and in the frequency spectra we find them around $30-300$~Hz. The STAFF instrument of the Cluster mission   \cite{Cornilleau2003} can in principle cover such frequencies. However, the turbulence level at $100-300$~Hz is below $\sim 10^{-7}$~nT$^2$/Hz, i.e., very close to the instrument noise level.   We will analyze in details the instrumental noise and its influence on the turbulent spectrum. 

In our analysis we use measurements of the STAFF-Spectrum Analyser (SA) which provides four seconds averages of the power spectral density (PSD) of the magnetic fluctuations at 27 logarithmically spaced frequencies, between 8~Hz and 4~kHz. 

\begin{figure}
\includegraphics[width=6.5cm]{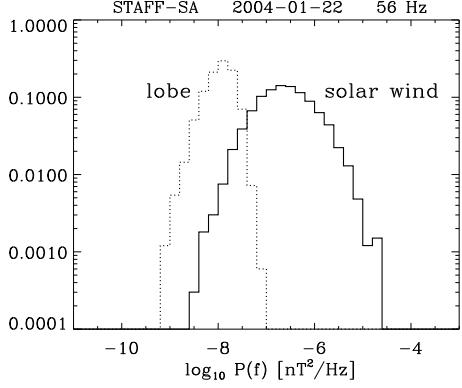}
\caption{\label{fig:histo-sa} Histograms of the total PSD of magnetic fluctuations measured by STAFF-SA/Cluster instrument at $f=56$~Hz in the lobe (left) and in the solar wind (right) during intervals of one hour. }
\end{figure}

To estimate the noise of STAFF-SA instrument, we use measurements  in the magnetospheric lobes. In this region the natural magnetic activity is negligible and  it may safely be assumed that what is measured is only the instrumental noise \cite{Cornilleau2003}. We use lobe data on the 27 January 2004, between 19:00 and 20:00~UT.  The measured power spectral density of a random noise is itself a random variable. The left (dotted) histogram of Figure~\ref{fig:histo-sa} shows that the distribution  $H(p_{lobe})$ of the total power spectral density in the lobe, $p_{lobe}$,  at a fixed frequency 56~Hz has a log-normal distribution, that is compatible with a multiplicative noise [P. Kellogg, private communication]. The expectation value  of this distribution is $\langle P_{noise}\rangle=\int p_{lobe} H(p_{lobe})dp_{lobe}=1.3\times 10^{-8}$nT$^2$/Hz. At the other frequencies of STAFF-SA we observe similar histograms, but with different mean PSD. The right histogram shows the distribution $H(p_{sw})$ of the same quantity and for the same frequency 56~Hz but measured during one hour in the free solar wind, with a rather important level of turbulence. The expectation value of  this last histogram is $\langle P_{sw}\rangle =\int p_{sw} H(p_{sw})dp_{sw} = 6.1 \times 10^{-7}$nT$^2$/Hz. One can see that even if the expectation value  of the solar wind distribution is much higher than  the one of the lobe distribution, there is an intersection of the histograms, and the measurements of the mean solar wind turbulence energy can be affected by the instrumental noise. Let us now discuss this aspect of the measurement in more details.

\begin{figure}
\includegraphics[width=6.5cm]{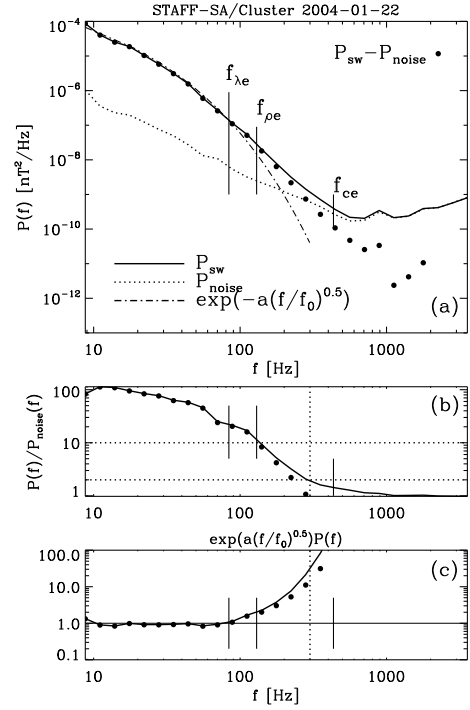}
\caption{\label{fig:spec-sa} (a) Solid line shows the spectrum in the solar wind  measured by STAFF-SA/Cluster instrument on 22 January 2004, 05:03-05:45UT. Dotted line shows the instrumental noise level. Black dots show the corrected spectrum $P_{turb}=(P_{sw}-P_{noise})$. Dashed-dotted line indicates exponential fit $\sim \exp (-a(f/f_0)^{0.5})$, with $f_0=f_{\rho_{e}}=V/2\pi\rho_e$ and the constant $a\simeq 9$. (b) Ratios $P_{sw}/P_{noise}$ (solid line) and $P_{turb}/P_{noise}$ (dots); horizontal lines indicate values  2 and 10.   
(c)  Spectra  $P_{sw}$ (solid line) and $P_{turb}$ (black dots) compensated by  the exponential. In all the panels, vertical solid bars indicate electron characteristic scales. 
In panels (b) and (c) vertical dotted line at 300~Hz indicates the maximal frequency of our analysis.}
\end{figure}

The distribution $H(p_{sw})$ is a superposition of the distribution of the turbulent signal  $H(p_{turb})$ and of the noise one $H(p_{lobe})$.  It is well known that the expectation value of the sum of two random variables is the sum of the expectation values of these variables. Therefore, supposing the independence of the noise from the physical signal, the mean PSD of the turbulent signal at each frequency can be determined as the difference between the corresponding expectation values 
\begin{equation}
\langle P_{turb} \rangle = \langle P_{sw}  \rangle  - \langle P_{noise}  \rangle. 
\end{equation}
In Figure~\ref{fig:spec-sa}(a) the solid line represents the mean solar wind spectrum $\langle P_{sw}  \rangle$ from 8~Hz to 4~kHz, the thin dotted line indicates  $\langle P_{noise} \rangle$ and black dots show the resulting turbulent spectrum  $\langle P_{turb} \rangle$ (in the  legend and caption of the plot, the brackets $\langle \cdot \rangle$ are omitted). Panel (b) shows the ratios $\langle P_{sw}  \rangle / \langle P_{noise} \rangle$ (solid line) and   $\langle P_{turb}  \rangle / \langle P_{noise} \rangle$ (dots).  From these two upper panels of Figure~\ref{fig:spec-sa}, one can see that for $f\ge 10^3$~Hz, the solar wind spectrum is identical with the noise spectrum, here $\langle P_{sw}  \rangle/ \langle P_{noise} \rangle = 1$.  At $f<10^3$~Hz, $\langle P_{sw} \rangle$ is above the noise; however, already at $140$~Hz, where $\langle P_{sw}  \rangle/ \langle P_{noise} \rangle = 10$, $\langle P_{sw} \rangle$ is affected by the noise, as it starts to deviate from  $\langle P_{turb} \rangle$. The noise becomes important for $f \ge 300$~Hz, where $\langle P_{sw}  \rangle/ \langle P_{noise} \rangle \le 2$.   Thus, we cannot just use the measured spectrum  up to the frequency where its level meets the noise level, i.e. $\langle P_{sw}  \rangle/ \langle P_{noise} \rangle = 1$, as the solar wind turbulent spectrum;  we need to take into account the effect  of the instrumental noise.  A meaningful solar wind turbulence spectrum is the corrected spectrum when it remains above the noise. So, in our particular case of Figure~\ref{fig:spec-sa}, the maximal frequency is $300$~Hz (see the vertical dotted line in Figure~\ref{fig:spec-sa}(b) and (c)). 

Now, let us focus on the spectral shape. The dashed-dotted line in Figure~\ref{fig:spec-sa}(a) gives an exponential fit $\sim \exp (-a\sqrt{f/f_{\rho_e}})$. This is the best fit with $\langle P_{sw} \rangle$ and $\langle P_{turb} \rangle$ in the frequency range $[8,100]$~Hz, where the noise doesn't affect the observed spectrum and both spectra are identical. The important result is that the spectrum is no more a power-law but curved over a broad frequency range  $\sim[10,100]$~Hz: this implies a deviation from self-similarity and the possible role of a dissipative process competing with non-linearities. The exponential fit can seem to be arbitrary. However, the compensated spectra shown in Figure~\ref{fig:spec-sa}(c) are indeed very flat for the whole decade: that confirms the exponential spectral shape.

\section{Spectrum {\bf at MHD and ion scales}}

Now, let us consider the combination of STAFF-SA spectra with spectra measured  by FGM \cite{Balogh2001} and STAFF-SC \cite{Cornilleau2003} instruments at lower frequencies. We have analyzed such combined spectra in \cite{alex09}. 
It was shown that for a quasi-perpendicular configuration between the mean solar wind velocity ${\bf V}$ and the magnetic field ${\bf B}$, under different plasma conditions, magnetic spectra show a quasi-universal form: $\sim f^{-5/3}$--power law at MHD scales and   $\sim f^{-2.8}$--power law at ion scales. 
At electron scales, the spectrum is no more a power-law but is curved, as we confirm here with Figure~\ref{fig:spec-sa}. 

\begin{figure}
\includegraphics[width=5.5cm]{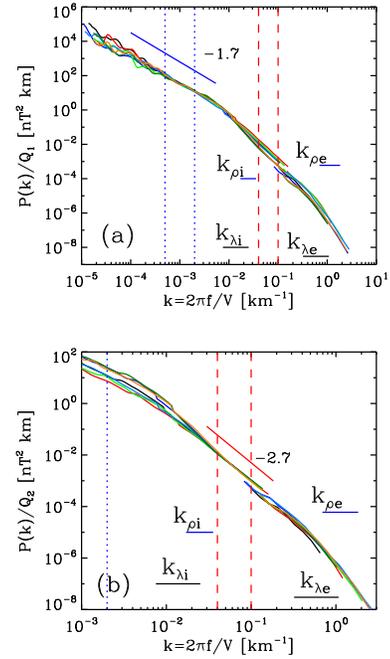}
\caption{\label{fig:Q1_Q2} (a) Spectra for 7 time intervals of 42 minutes in the solar wind, studied in \cite{alex09}, rescaled by $Q_1$; (b) The same spectra as in (a), but for $10^{-3}<k<3$~km$^{-1}$, rescaled by $Q_2$. Vertical dotted (dashed) lines indicate wave vector range where $Q_1$ ($Q_2$) is determined;  solid lines refer to the power laws $k^{-1.7}$ and $k^{-2.7}$; horizontal bars indicate plasma characteristic scales. Color code corresponds to different time periods, listed in \cite{alex09}.} 
\end{figure}

This quasi-universal spectrum was obtained by a superposition of seven spectra using (i) the Taylor hypothesis $k=2\pi f/V$, $P(k)=P(f)V/2\pi$ and (ii) an intensity factor $Q_0(j)=\langle P_j(k)/P_1(k) \rangle$, where $P_1(k)$ is a reference spectrum and $\langle \cdot \rangle$ a mean over the range of wave vectors covering MHD and ion scales.  
Let us now superpose the spectra independently at MHD scales and at ion scales. For this we define two different factors, $Q_1$ and $Q_2$.  $Q_1$ is determined in an interval of the inertial range $[0.5,2]\times 10^{-3}$~km$^{-1}$ indicated by vertical dotted lines in  Figure~\ref{fig:Q1_Q2}(a). Here, the spectra are clearly superposed, following a $k^{-1.7}$ power law.  At higher $k$, the spectra first spread  just above the break point,  around the spatial ion scales $k_{\lambda_i}=1/\lambda_i$, $k_{\rho_i}= 1/\rho_i$, and then, starting at $k\simeq 0.04$~km$^{-1}$, the spectra  appear to be parallel to each other again.  In the range where the spectra are parallel, we chose an interval $k=[0.04,0.1]$~km$^{-1}$ (between vertical dashed lines) to define $Q_2$. Normalization on $Q_2$ gives us Figure~\ref{fig:Q1_Q2}(b), where the seven spectra are superposed perfectly  and form one clear power law $\sim k^{-2.7}$, with a very small spectral index  dispersion  of $0.1$.   That is close to the results obtained in \cite{alex09},  but more precise.

The analysis presented in Figure~\ref{fig:Q1_Q2} shows that turbulent spectrum at MHD and ion scales can 
be characterized within three spectral ranges:  (1) a quasi-universal power-law $\sim k^{-5/3}$ at MHD scales, 
(2) spectral spread around the break point and (3) a quasi-universal spectrum  $\sim k^{-2.7}$ at $k>k_{\rho_i}$. 
Note that the spectral range  (2) corresponds  to  $[0.5,2]$~Hz, approximately the range where Smith  {\it et al.} \cite{Smith2006}  observe the dispersion of the spectral index as well.  The spectral range (3) corresponds to $f\simeq [2,10]$~Hz, and it has not been observed before because of the white noise at FGM instruments.  Here we observe it thanks to a high sensitivity of the STAFF-SC instrument at these frequencies.

It was recently shown that ion instabilities observed in the solar wind \cite{Matteini07} can locally generate magnetic fluctuations at scales close to the break of the solar wind turbulent spectrum \cite{Bale09}. It will be interesting to understand the role of these instabilities in the variability of the spectra around the break.

\section{Dimensionless spectra}

\begin{figure}
\includegraphics[width=7.5cm]{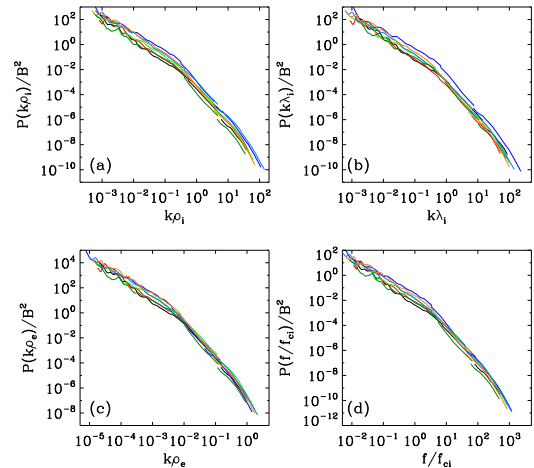}
\caption{\label{fig:dimless} Normalized dimensionless spectra as  functions of (a) $k\rho_i$, (b) $k\lambda_i$, (c) $k\rho_e$ and (d) $f/f_{ci}$.} 
\end{figure}

It is interesting to compare turbulent spectra under different plasma conditions not only at the same 
 $k$ in km$^{-1}$, as we did in the previous section, but at the same $k$ or $f$ normalized with the characteristic plasma scales, such as $\rho_{i,e}$, $\lambda_{i,e}$ and $f_{ci,e}$.  If $r$ is a characteristic plasma scale, we apply the following change of variables: 
\begin{equation}
k \rightarrow kr, \;\; P(k) \rightarrow P(kr)=P(k)\frac{1}{r}.
\end{equation}
Such spectra have the dimension of $nT^2$.  Normalization over $B^2$ yields  dimensionless spectra presented in Figure~\ref{fig:dimless}. Here panel (a) shows the normalized spectra as a function of $k\rho_i$, (b) $k\lambda_i$, (c) $k\rho_e$ and (d) $f/f_{ci}$.  

An advantage of this representation is that such turbulent spectra in the solar wind can be directly compared with any magnetic spectrum of different  astrophysical or plasma device turbulent systems and without any assumptions on turbulent models. 

It is a long standing problem to distinguish between different plasma scales: which of them is responsible for the spectral break at ion scales and which of them plays the role of the dissipation scale in space plasmas. From Figure~\ref{fig:dimless}(a) and (b) it is still difficult to say which of the ion scales is responsible for the spectral break: the break is observed closer to $k\lambda_i =1$ than to $k\rho_i =1$, but this is not enough to make a conclusion.   It is possible that both scales are crucial for the change of the turbulence nature at the limit of the MHD description [M. Velli, private communication].

Now, let us consider the $k\rho_e$--normalization, Figure~\ref{fig:dimless}(c). All the spectra nearly collapse at the spectral break at ion scales ($k\rho_e\simeq 10^{-2}$), and at higher $k\rho_e$ the spectra are very close to each other. This distinguishes electron gyro-radius from the other spatial plasma scales.  A similar but less clear collapse is observed in panel (d), where the spectra $P(f/f_{ci})/B^2$ are shown. These observations confirm our results presented in \cite{alex09}.

\section{Conclusions}

In this paper we analyze in details the problem of the noise of  the STAFF-SA instrument. We show that it influences the solar wind measurements when the signal to noise ratio is less than 10. 
The meaningful spectrum is the corrected spectrum where it remains above the noise. 

In the rest of the paper we  confirm our results presented in \cite{alex09}: (i) at ion scales for $f \simeq [2,10]$~Hz, the spectrum follows a quasi-universail power-law  $\sim k^{-2.7}$; (ii) in a broad range $\sim[10,100]$~Hz,  below $k\rho_e=1$,   spectrum is not a power-law but is exponential; (iii) for $k\rho_e>1$ (above 100 Hz), the spectrum deviates from the exponential, and it seems to follow a very steep power-law, but the spectral index is different for the measured and corrected spectrum. Therefore, to make any conclusion about the spectral shape at $k\rho_e>1$, observations in  regions with a higher signal to noise ratio, typically with $P_{sw}/P_{noise}>10$, are needed.

Finally,  we have discussed the role of different kinetic plasma scales on the spectral shape, considering normalized dimensionless spectra.

\bibliographystyle{aipproc}   

\bibliography{sw12-short}

\begin{thebibliography}{12}
\expandafter\ifx\csname natexlab\endcsname\relax\def\natexlab#1{#1}\fi
\providecommand{\enquote}[1]{``#1''}
\expandafter\ifx\csname url\endcsname\relax
  \def\url#1{\texttt{#1}}\fi
\expandafter\ifx\csname urlprefix\endcsname\relax\def\urlprefix{URL }\fi
\providecommand{\eprint}[2][]{\url{#2}}

\bibitem[Bruno and Carbone(2005)]{noi}
R.~Bruno, and V.~Carbone, \emph{Living Rev. in Sol. Phys.} \textbf{2}, 4
  (2005).

\bibitem[Leamon(1998)]{Leamon1998}
R.~J. Leamon, {\it et~al.}, \emph{J. Geophys. Res.} \textbf{103}, 4775 (1998).

\bibitem[Smith(2006)]{Smith2006}
C.~W. Smith, {\it et~al.}, \emph{Astrophys. J.} \textbf{645}, L85 (2006).

\bibitem[Alexandrova(2009)]{alex09}
O.~Alexandrova, {\it et~al.}, \emph{Phys. Rev. Lett.} \textbf{103}, 165003
  (2009).

\bibitem[Mangeney(2006)]{Mangeney2006}
A.~Mangeney, {\it et~al.}, \emph{Ann. Geophys.} \textbf{24}, 3507--3521 (2006).

\bibitem[Lacombe(2006)]{Lacombe2006}
C.~Lacombe, {\it et~al.}, \emph{Ann. Geophys.} \textbf{24}, 3523--3531 (2006).

\bibitem[Alexandrova(2008)]{alexandr08angeo}
O.~Alexandrova, {\it et~al.}, \emph{Ann. Geophys.} \textbf{26}, 3585 (2008).

\bibitem[Sahraoui(2009)]{Fouad09}
F.~Sahraoui, {\it et~al.}, \emph{Phys. Rev. Lett.} \textbf{102}, 231102 (2009).

\bibitem[Cornilleau-Wehrlin(2003)]{Cornilleau2003}
N.~Cornilleau-Wehrlin, {\it et~al.}, \emph{Ann. Geophys.} \textbf{21}, 437
  (2003).

\bibitem[Balogh(2001)]{Balogh2001}
A.~Balogh, {\it et~al.}, \emph{Ann. Geophys.} \textbf{19}, 1207 (2001).

\bibitem[Matteini(2007)]{Matteini07}
L.~Matteini, {\it et~al.}, \emph{Geophys. Res. Lett.} \textbf{34}, 20105
  (2007).

\bibitem[Bale(2009)]{Bale09}
S.~Bale, {\it et~al.}, \emph{Phys. Rev. Lett.} \textbf{103}, 211101 (2009).

\end{thebibliography}
\end{document}